\documentclass[aaspp4]{aastex}
\usepackage{psfig,epsf}

\shortauthors{J.J.M. in 't Zand et al.}
\shorttitle{BeppoSAX measurements of GRB~010222}

\begin{document}

\title{BEPPOSAX MEASUREMENTS OF THE BRIGHT $\gamma$-RAY BURST 010222}
\author{J.J.M.~in~'t~Zand\altaffilmark{1,2,3}, 
 L.~Kuiper\altaffilmark{2},
 L.~Amati\altaffilmark{4},
 L.A.~Antonelli\altaffilmark{5,6}
 R.C.~Butler\altaffilmark{4}
 E.~Costa\altaffilmark{7},
 M.~Feroci\altaffilmark{7},
 F.~Frontera\altaffilmark{4,8},
 G.~Gandolfi\altaffilmark{7},
 C.~Guidorzi\altaffilmark{8},
 J.~Heise\altaffilmark{2},
 R.G.~Kaptein\altaffilmark{9,2},
 E.~Kuulkers\altaffilmark{1,2},
 L.~Nicastro\altaffilmark{10},
 L.~Piro\altaffilmark{7},
 P.~Soffitta\altaffilmark{7},
 M.~Tavani\altaffilmark{11,12}}
\affil{}
\altaffiltext{1}{Astronomical Institute, Utrecht University, P.O. Box 80 000, 3508 TA Utrecht, the Netherlands}
\altaffiltext{2}{Space Research Organization Netherlands, 
	Sorbonnelaan 2, 3584 CA Utrecht, the Netherlands}
\altaffiltext{3}{email jeanz@sron.nl}
\altaffiltext{4}{Istituto di Tecnologie e Studio delle Radiazioni Extraterrestri (CNR), Via Gobetti 101, 40129 Bologna, Italy}
\altaffiltext{5}{Osservatorio Astronomico di Roma, Via Frascati 33, 00040 Monteporzio Catone, Italy}
\altaffiltext{6}{BeppoSAX Science Data Center, Via Corcolle 19, 00131 Rome, Italy}
\altaffiltext{7}{Istituto di Astrofisica Spaziale (CNR), 00133 Rome, Italy}
\altaffiltext{8}{Dipartimento Fisica, Universit\'{a} di Ferrara, Via Paradiso 12, 44100 Ferrara, Italy}
\altaffiltext{9}{BeppoSAX Scientific Operation Center, Via Corcolle 19, 00131 Rome, Italy}
\altaffiltext{10}{Istituto Fisica Cosmica e Applicazioni all'Informatica (CNR), Via Ugo La Malfa 153, 90146 Palermo, Italy}
\altaffiltext{11}{Istituto Fisica Cosmica e Tecnologie Relative (CNR), Via Bassini 15, 20133 Milan, Italy}
\altaffiltext{12}{Columbia Astrophysics Laboratory, Columbia University, New York, NY 10027, U.S.A.}

\begin{abstract}
We analyze the BeppoSAX measurements of the prompt and afterglow emission of
the $\gamma$-ray burst GRB~010222. Among 45 GRBs detected with the Wide
Field Cameras on BeppoSAX, the 40-700 keV fluence of
$(9.3\pm0.3)\times10^{-5}$~erg~cm$^{-2}$ is only surpassed by GRB~990123. In
terms of the isotropic 20-2000~keV energy output of $7.8\times10^{53}$~erg,
it ranks third of all GRBs with measured distances.
Since this burst is so bright, the data provide complete and valuable
coverage up to 65~hr after the event, except for a gap between 3.5 and 8.0~hr. 
The 2-10 keV flux history shows clear signs of a break which is consistent
with a break seen in the optical, and provides supporting evidence for the
achromatic nature of the break. An explanation for the break in the context of
a collimated expansion is not straightforward. Rather, a model is favored
whereby the fireball is braked to the non-relativistic regime quickly (within
a fraction of day) by a dense ($\sim10^6$~cm$^{-3}$) circumburst medium. 
This implies that, after a mild beaming correction, GRB~010222 may be the most
energetic burst observed thus far. The X-ray decay index after the break
is $1.33\pm0.04$, the spectral index $0.97\pm0.05$. The decay is,
with unprecedented accuracy, identical to that observed in the optical.
\end{abstract}



\section{Introduction}
\label{secintro}

In its nearly five year mission up to March 2001, the Wide Field Cameras
(WFCs) and Gamma-Ray Burst Monitor (GRBM) on the BeppoSAX platform
simultaneously detected 45 $\gamma$-ray bursts (GRBs) that were analyzed
in near-real time. An increasing number of bursts are sampled
at the high end of the peak flux distribution. In this paper we discuss
GRB~010222 which is the most energetic burst detected after GRB~990123.

The GRBM was triggered by GRB~010222
on 2001, Feb.22.3073484 U.T. Simultaneously, WFC unit 1 detected this burst
at an off-axis angle of 15$^{\rm o}$. The WFC detection is of high statistical
significance and the burst could be localized within 2\farcm5 (Piro 2001a). 
An alert message was distributed 3.2 hr after the burst, and follow-up
studies were quick thanks to the favorable declination and timing of the burst.
A $V=18$ optical counterpart was publicly announced within 4.4~hr after the
burst (Henden 2001a, 2001b); a radio counterpart at 22~GHz within 7.7~hr
(Berger \& Frail 2001), and detections followed in the $R$-band (Stanek et
al. 2001a), near infrared (Di Paola et al. 2001) and at sub-mm wavelengths
(Fich et al. 2001). Remarkably, the first report of a redshift
appeared within only 11.4~hr (Garnavich et al. 2001a). Later studies
(Jha et al. 2001a, Bloom et al. 2001a, Jha et al. 2001b) revealed
absorption systems at three different redshifts, with $z=1.477$ the largest.
Jha et al. (2001b) argue that the $z=1.477$ system is actually from the host
galaxy.

We here present all GRB~010222 measurements that were obtained with BeppoSAX
instruments. These pertain to the burst itself (in X-rays and
$\gamma$-rays) and the X-ray afterglow. Since the burst is so bright, the
statistical quality of these data is high and allows for a sensitive study of
various aspects of this burst. We discuss the radiation output
of the burst and the evidence for a dense circumburst environment.

\section{The burst event}
\label{sectionburst}

The burst event was measured with two instruments on board BeppoSAX
(Boella et al. 1997b). The GRBM (Amati et al. 1997 and Feroci et al. 1997) 
comprises 4 lateral shields of the Phoswich Detector System (PDS, Frontera
et al. 1997) and has a bandpass of 40 to 700 keV. The normal directions
of two shields are each co-aligned with the viewing direction of a WFC unit.
The WFCs (Jager et al. 1997) consist of two identical coded aperture cameras 
each with a field of view of 40$^{\rm o}\times40^{\rm o}$ full-width to zero 
response and an angular resolution of about 5\arcmin. The bandpass is 2 to 28 
keV. 

\begin{figure}[t]
  \begin{center}
    \leavevmode
\epsfxsize=13cm
\epsfbox{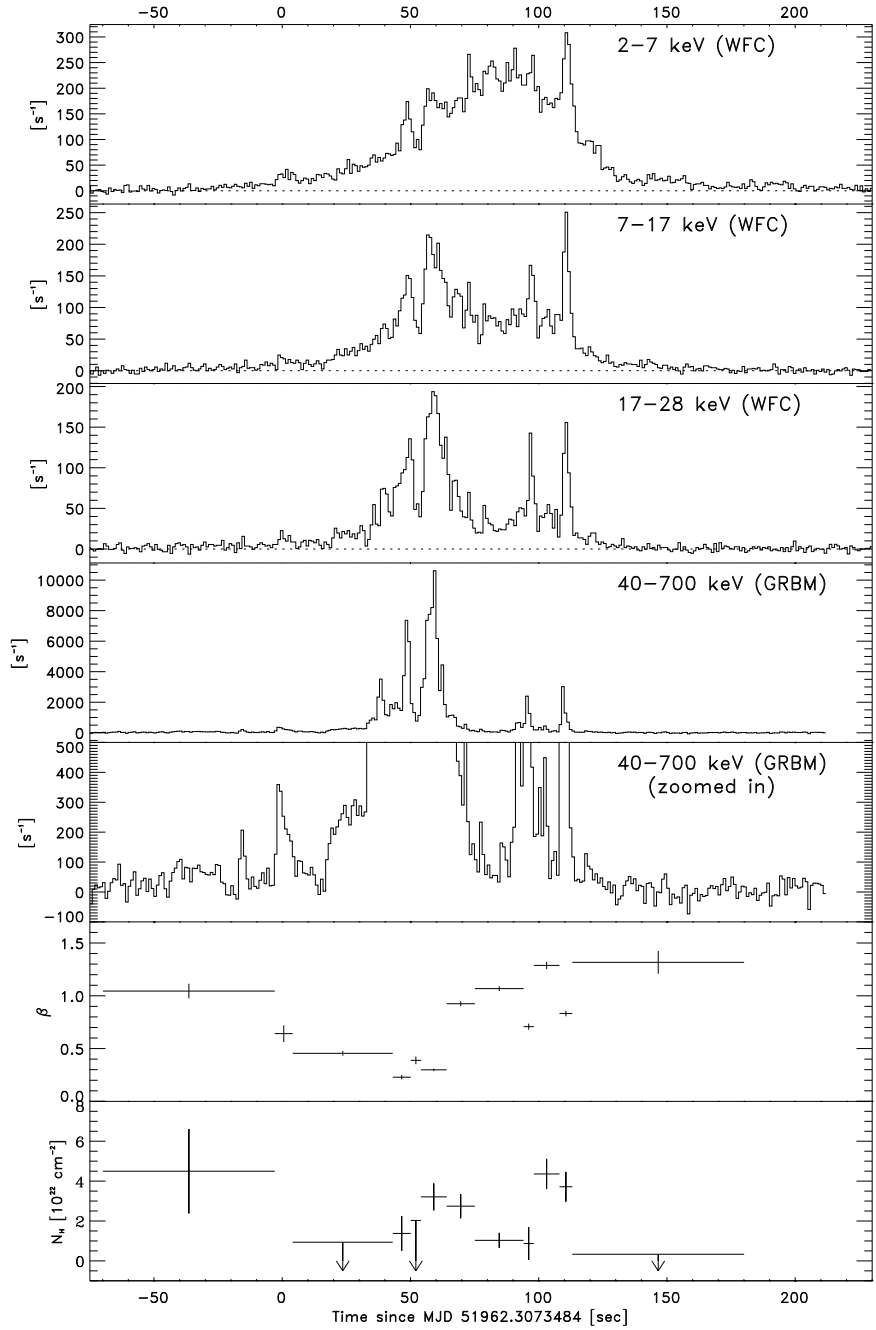}
  \caption{Time history of the burst itself as seen with WFC and GRBM, 
at a time resolution of 1~s. The two lower panels present the evolution
of the power
law index $\beta$ and absorption column $N_{\rm H}$ as simultaneously
determined from WFC and 2-channel GRBM data, except for the last data
point which did not contain a significant GRBM signal. Upper limits
for $N_{\rm H}$ are $1\sigma$ values.\label{figpromptlc}
}
  \end{center}
\end{figure}

In figure~\ref{figpromptlc}, the time profile of the burst is shown in various
bandpasses. The $\gamma$-ray light curve consists of seven clear pulses over
a $\approx$130~s time interval, with recursions to almost the quiescent
level in between. The GRBM triggered on the second pulse. At a resolution of
7.8125~ms (not shown) the main peak shows 5 sub-pulses. The peak flux is
30$\pm$5\% higher at 62.5~ms than at 1~s time resolution. The Fourier power
spectrum shows no narrow features. When carefully investigating the
$\gamma$-ray flux before the first pulse (5th panel in
figure~\ref{figpromptlc}), it shows a slight increase at 40~s before the
trigger time, with rates going up to about 100~s$^{-1}$ (or 1\% of the peak
rate). The X-ray light curve starts off at approximately the same moment. It
is characterized by a slow rise which persists during the first $\gamma$-ray
pulses. This behavior is reminiscent of other bursts, some of which have
apparent X-ray precursor activity (e.g., In 't Zand et al. 1999). Perhaps
in some of those cases we are simply missing the weak $\gamma$-ray emission
because the flux does not exceed the detection threshold.
Noteworthy is also that the X-radiation continues for some
150~s after the cessation of the $\gamma$-radiation.
Both the X-ray and $\gamma$-ray light curves show strong spectral evolution.

In~table~\ref{table1} we list the duration, peak flux and fluence of the
prompt emission.
In many respects, GRB~010222 ranks in the top three of GRBs that were detected
with WFC. In terms of 2-28 keV peak flux, GRB~010222 ranks second after
GRB~990712 (by a factor of 1.8; Frontera et al. 2001).
The X-ray duration of $\approx$280~s is among the longest measured with WFC,
together to GRB~980519 (250 s,
In~'t~Zand et al. 1999), GRB~981226 (250 s, Frontera et al. 2000) and
GRB~990907 (230 s). The combination of longevity and brightness makes this
burst number one in X-ray fluence: its fluence is 3 times larger than
the next burst in line (GRB~990712). In 40-700 keV peak flux, it
ranks third after GRB~000210 (Stornelli et al. 2000) and GRB~990123
(Feroci et al. 1999), and in fluence it
ranks second after GRB~990123. The 50-300 keV
burst fluence places it in the top 1.3\% of the BATSE burst sample and 
the 50-300 keV peak photon flux in the top 2.5\%.

\begin{figure}[t]
  \begin{center}
    \leavevmode
\epsfxsize=13cm
\epsfbox{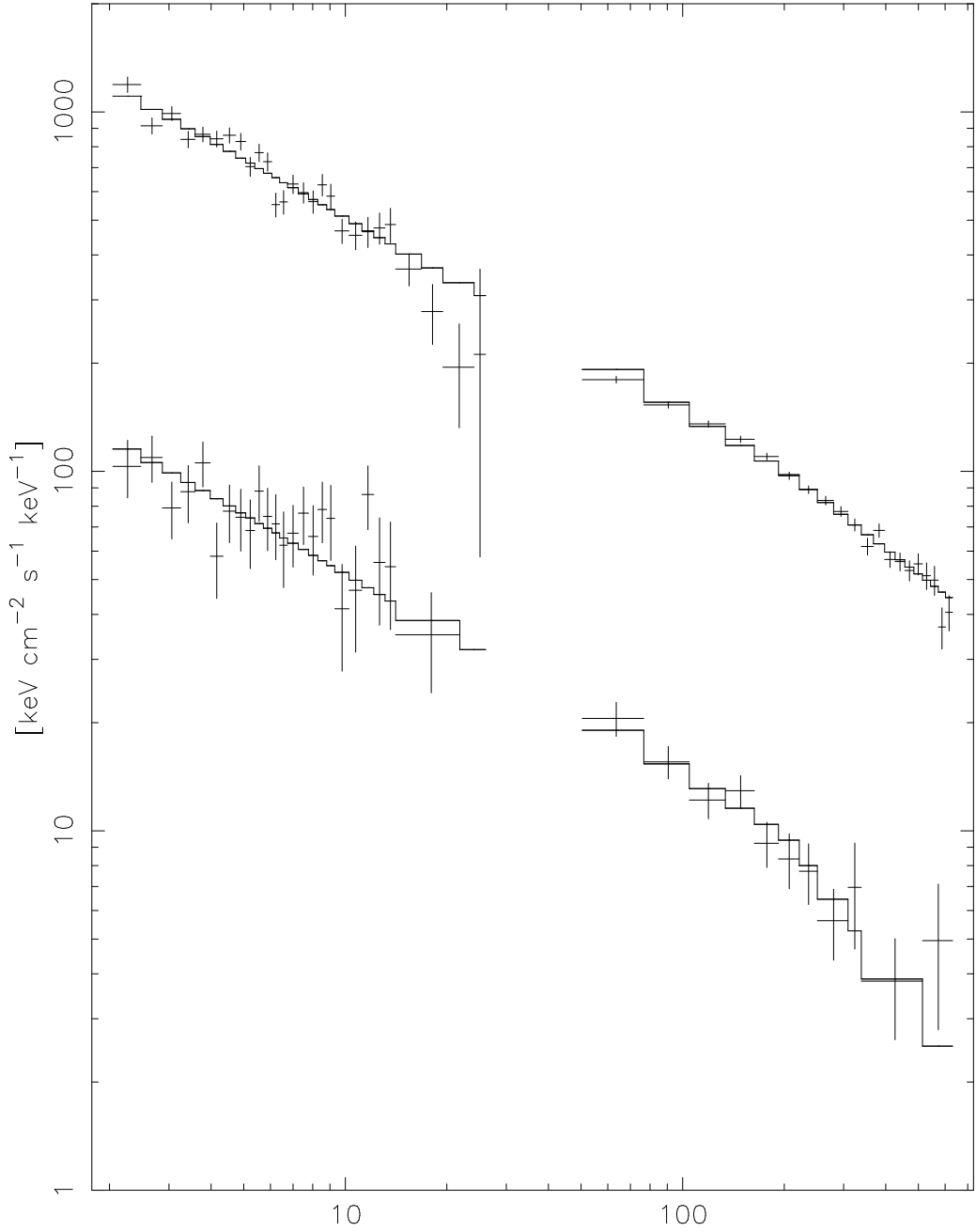}
  \caption{Combined WFC and GRBM spectra for intervals -91/+37 (lower
curve) and +37/+165~s (upper curve) after the trigger time. The solid
lines indicate the best-fit absorbed broken power law function with the same
absorption and break energy in both cases. The $\chi^2_{\nu}=1.346$ for
$\nu=88$ dof, the break energy is $210\pm30$~keV. There is no improvement
in the fit if the break energy is left free in each interval. The 
spectral index $\beta$ is $0.56\pm0.02$ below and 
$1.3\pm0.3$ above the break for the lower curve and $0.543\pm0.007$ and
$0.72\pm0.04$ respectively for the upper curve.
\label{figlongspectra}
}
  \end{center}
\end{figure}

The GRBM continuously samples 256-channel spectra every 128~s between 40
and 700 keV. 240 of these channels are well calibrated, roughly between 50 and
650~keV. The phasing of the accumulation timing is arbitrary. GRB~010222 is
covered by two 128~s accumulation intervals that meet at 37~s after the trigger
time. In figure~\ref{figlongspectra} we show these spectra, combined with the
appropriate WFC data. There is evidence for a break at 210 keV and for a
low-energy cutoff that may be fitted with absorption due to cold interstellar
matter. 
The GRBM also provides photon rates at 1-s resolution in 40-700 keV and
$>100$~keV. 
If these data and the WFC data are resolved in 12 time intervals
(see bottom panels of figure~\ref{figpromptlc}), and an absorbed power
law is fitted\footnote{formulated by $F(E) \sim {\rm exp}(-N_{\rm H}\sigma(E))
E^{-\beta}$~keV~s$^{-1}$cm$^{-2}$keV$^{-1}$,
where $\sigma(E)$ is the cross section as a function of photon energy $E$
for cold interstellar matter of cosmic abundances according to the model by
Morrison \& McCammon (1983), $N_{\rm H}$ is the hydrogen column
density and $\beta$ is the energy index}
where the index is left free over all 12 intervals while $N_{\rm H}$ is a
single free parameter over all intervals, $\beta$ varies between 0.0 and
1.5, while $N_{\rm H}=(1.7\pm0.2)\times10^{22}$~cm$^{-2}$. The values for
$\beta$ are in good agreement with the shock synchroton model put forward
by Tavani (1996). If $N_{\rm H}$
is left free, $\chi^2_\nu$ improves significantly from 0.943 (358 dof) to
0.931 (348 dof). An f-test shows a negligible chance probability.
$N_{\rm H}$ varies between non-detections with 3$\sigma$ upper limits as small
as $1.0\times10^{22}$~cm$^{-2}$ (for the last interval) to a peak of
$4\times10^{22}$~cm$^{-2}$
(see figure~\ref{figpromptlc}). Note that $N_{\rm H}$ for the Galactic
absorption in the direction of the burst is $1.6\times10^{20}$~cm$^{-2}$
according to an interpolation of the maps published by Dickey \& Lockman
(1990). 

The WFC observation on this field was from 17.8~hr before the burst till
3.5~hr afterwards. A total of 16.9~ks of net exposure time was accumulated.
We could not find any signal from the burst position on time scales of 5~s,
10~s, 1000~s or 3.4~hr, and in 2-10 or 2-28 keV. The 3$\sigma$ upper limit
for the remainder of the observation, starting at 260~s after the trigger time,
is $1.7\times10^{-10}$~erg~s$^{-1}$cm$^{-2}$ (2-10 keV; 3.6~ksec net
exposure time). This is a remarkable upper limit, see Sect.~\ref{sectiondis}.
The 3$\sigma$ upper limit for all data obtained prior to 70~s before the
trigger time is $9\times10^{-11}$~erg~s$^{-1}$cm$^{-2}$ (13.0 ksec net exposure
time).

\section{The X-ray afterglow}
\label{xrayafterglow}

\begin{figure}[t]
  \begin{center}
    \leavevmode
\epsfxsize=13cm
\epsfbox{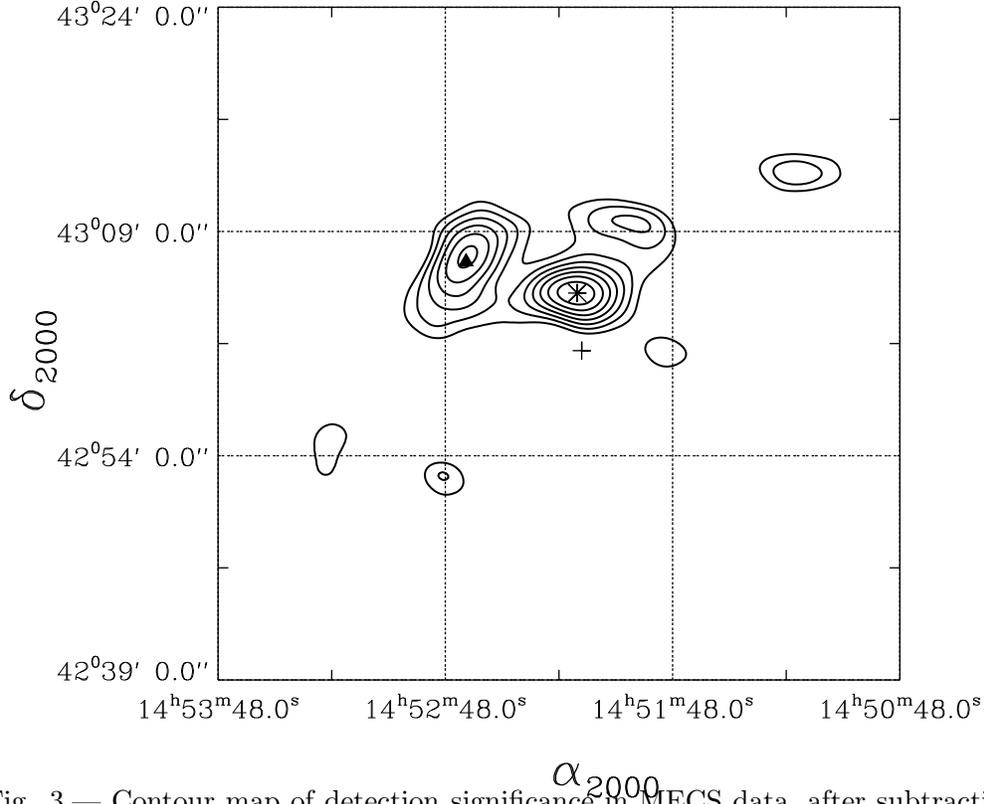}
  \caption{Contour map of detection significance in MECS data, after
subtraction of the brightest source (the cross labels the best-fit position).
Contours start at 4$\sigma$ and have steps of 1$\sigma$ for 1 d.o.f.
Three other sources are identified within 10\arcmin\ from the
afterglow, the two brightest are taken into account when generating
light curves (figure~\ref{fignfilc}) and spectra (figure~\ref{fignfispec}).
The best-fit positions of these are labeled with an asterisk and a triangle.
Their positions are $\alpha_{2000}~=~14^{\rm h}52^{\rm m}13.2^{\rm s}$,
$\delta_{2000}~=~+43^{\rm o}04\arcmin52\farcs8$ and
$\alpha_{2000}~=~14^{\rm h}52^{\rm m}42.5^{\rm s}$,
$\delta_{2000}~=~+43^{\rm o}07\arcmin7\farcs5$.
A similar map of the LECS data (not shown) reveals no sources other than
the afterglow.
\label{figmecsmap}
}
  \end{center}
\end{figure}

The BeppoSAX Narrow-Field Instruments (NFI) consist of four devices, among them
the Low-Energy (0.1 to 10 keV) and the Medium-Energy (2 to 10 keV) Concentrator
Spectrometer (LECS and MECS respectively, see Parmar et al. 1997 and Boella et
al. 1997a respectively). The NFI followed up on GRB~010222 from February 22.68
(8.0~h after the burst) to February 25.01 UT (65~hr after the burst). 
The MECS was not turned on until 1.3~hr after the start. The
net exposure time for the MECS is 88.4~ks, and for the LECS 50.3~ks.
The full-bandpass spatial MECS data
were studied applying a maximum likelihood method (e.g., In~'t~Zand
et al. 2000). Significant detections of four point sources were
identified within 10\arcmin\ from the WFC position (figure~\ref{figmecsmap}).
A light curve of the brightest source (figure~\ref{fignfilc}) shows a
typical decay and unambiguously identifies the source with the
X-ray afterglow of GRB~010222. The decay is consistent with a power-law
function $t^{-\alpha}$ ($t$ time since GRB) with $\alpha=1.33\pm0.04$
($\chi^2_\nu=1.01$ for $\nu=31$ dof). Such a decay index is very common
for an X-ray afterglow. The flux at 8.0~hr of
$1.2\times10^{-11}$~erg~s$^{-1}$cm$^{-2}$ is the brightest flux detected from
any X-ray afterglow at the same epoch, except for GRB~991216 which was 3
times brighter (Takeshima et al. 1999). The position of the X-ray source is
$\alpha_{2000}~=~14^{\rm h}52^{\rm m}12.0^{\rm s}$,
$\delta_{2000}~=~+43^{\rm o}01\arcmin01\farcs6$ (error radius
30\arcsec\ at 90\% confidence). This is 7\arcsec\ from the X-ray
position determined with the Chandra X-ray Observatory CXO (Garmire et al.
2001) and 8\arcsec\ from the optical transient (Henden 2001c, McDowell
et al. 2001)

\begin{figure}[t]
  \begin{center}
    \leavevmode
\epsfxsize=10cm
\epsfbox{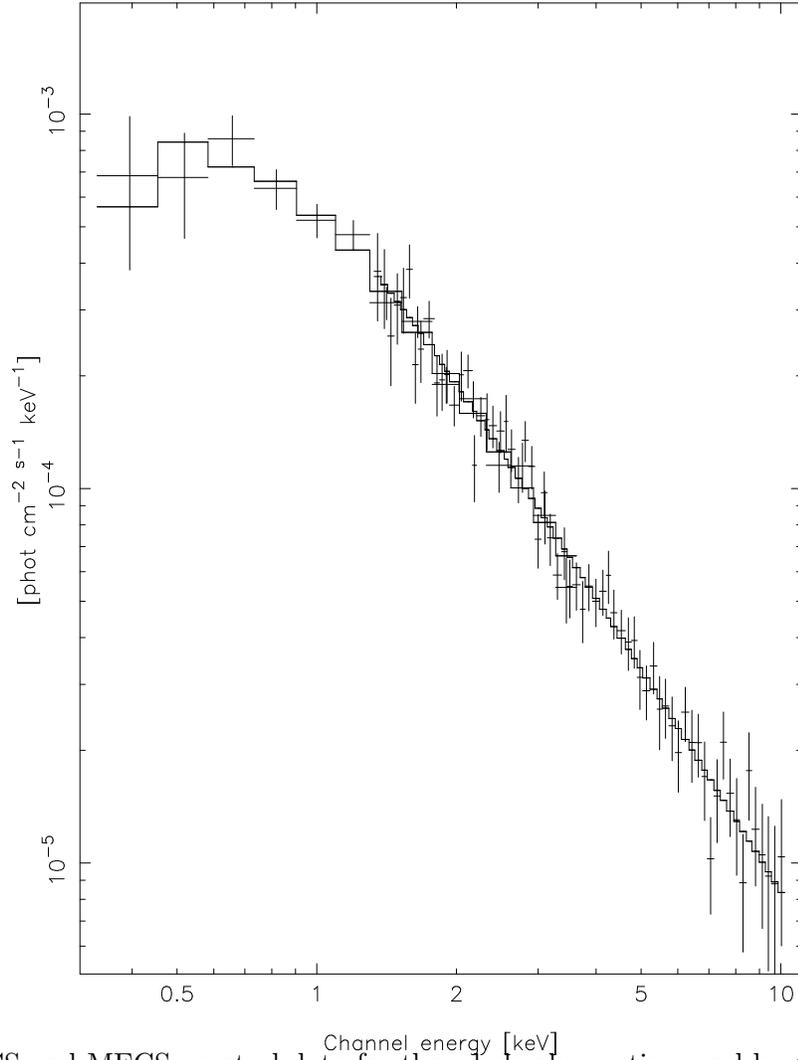}
  \caption{LECS and MECS spectral data for the whole observation, and best fit
absorbed power-law function. The data were extracted through a
maximum-likelihood fit to all point sources. This method retrieves all
detected source photons. The LECS to MECS normalization was allowed to
vary and converged to $0.97\pm0.05$.
\label{fignfispec}
}
  \end{center}
\end{figure}

The overall energy spectrum could be modeled with an absorbed power law, with
spectral index $\beta=0.97\pm0.05$ (again a very common value among X-ray
afterglows) and $N_{\rm H}=(1.5\pm0.3)~10^{21}$~cm$^{-2}$ ($\chi^2_{\nu}=0.83$
for $\nu=74$ dof, see figure~\ref{fignfispec}). $N_{\rm H}$ is one order of
magnitude smaller than during the prompt emission. Corrected for redshift,
$N_{\rm H}$ is $2.5\times10^{22}$~cm$^{-2}$.
The 2-10 keV fluence of the afterglow, integrated over the observation span
time, is 20 times smaller than that of the prompt emission. 
There is a small depression in the spectrum between
3 and 4 keV. We tested whether this feature is significant by modeling it
with an absorption edge which is parameterized by a threshold energy
$E_{\rm edge}$ and an optical depth $\tau$. We find that $\chi^2_{\nu}$
improves to 0.79 for $\nu=72$, with $E_{\rm edge}=3.1\pm0.2$~keV and
$\tau=0.19\pm0.10$. An f-test predicts that the probability for a chance
improvement is a marginal 0.7\%. When corrected for redshift, the edge energy
suggests K-edge absorption by Fe XIV-XVIII.

We searched the time-resolved and overall LECS+MECS spectrum for narrow
emission features by resolving the data logarithmically in 10 time bins,
and were unable to find conclusive evidence. The $3\sigma$ upper limit on a
narrow line at 6 to 7~keV in the overall spectrum is
$2\times10^{-6}$~phot~s$^{-1}$cm$^{-2}$, at
6.5/(1+$z$)=2.6~keV this is $1.2\times10^{-5}$~phot~s$^{-1}$cm$^{-2}$. 
For a Gaussian line with a width of 1~keV (FWHM) these upper limits are twice
as high. The upper limit for the redshifted narrow line is similar to the
flux of a $3\sigma$ detection of a narrow line at 4.7 keV in GRB~000214
(Antonelli et al. 2000). With respect to the broad emission line detected
in GRB~991216 at 3.5~keV (Piro et al. 2000), our upper limit is about two
times smaller.
We also studied, in the same 10 time bins, the 3.4-10 keV versus 1.8-3.4
keV hardness ratio and found no evidence for evolution in the continuum
shape.

\section{Discussion}
\label{sectiondis}

For a redshift of $z=1.477$ (Jha et al. 2001b) and a
\{$H_0$=65~km~s$^{-1}$Mpc$^{-1}$,
$\Omega_{\rm m}=0.3$, $\Lambda=0.7$\} cosmology, the luminosity distance is
$D_{\rm L}=3.6\times10^{28}$~cm. If the emission is assumed to be isotropic,
the $\gamma$-ray energy output is
$E_{\rm iso}(\gamma)=4\pi D_{\rm L}^2F_\gamma/(1+z)$ where $F_\gamma$ is the
observed fluence. In the 2 to 700 keV bandpass, this is $7.7\times10^{53}$~erg.
In the redshift-corrected (or 'co-moving', see Bloom, Frail \& Sari 2001b)
20-2000~keV bandpass this is $7.8\times10^{53}$~erg. These
two numbers are so close because the 20-2000~keV co-moving
band is close the instrument band for the observer (8-807 keV).
When compared to the 17 GRBs with established redshifts prior to
GRB~010222 (Frail et al. 2001, Bloom et al. 2001b), there are only
two other GRBs with higher values for $E_{\rm iso}(\gamma)$ in the
co-moving 20-2000 keV band: GRB~990123
(by a factor of 1.8) and GRB~000131 (by a factor of 1.5; this burst was not
visible to BeppoSAX). 

\begin{figure}[t]
  \begin{center}
    \leavevmode
\epsfxsize=13cm
\epsfbox{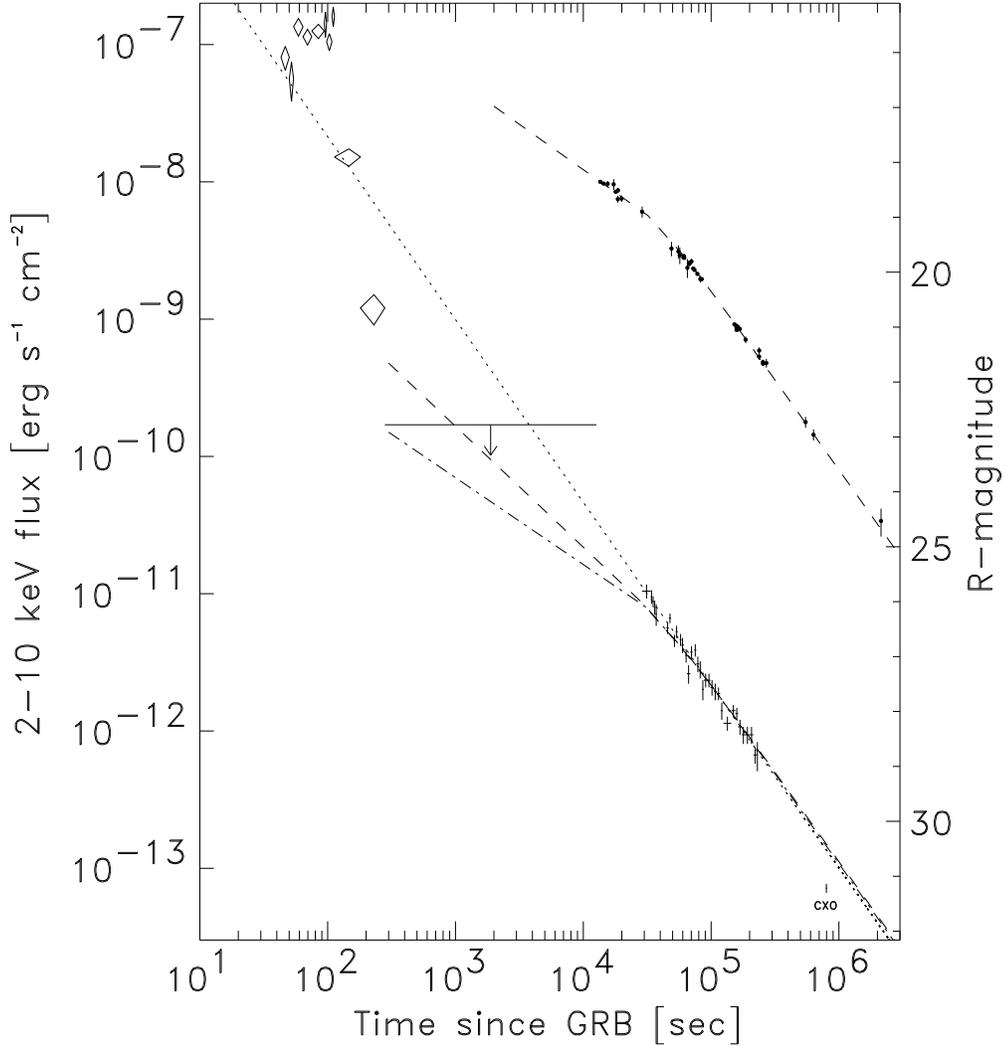}
  \caption{2-10 keV flux history (left odinate) of the prompt emission
(sole upper limit and diamonds whose sizes represent error margins) and
afterglow emission (crosses, based on MECS data, except for the first data
point which is determined from LECS data and the last data point which is based
on CXO data by Harrison et al. 2001b). A constant
spectral shape was assumed throughout each data set to translate photon count
rates to  energy flux. The filled circles in top right-hand corner represent
photometry in the Cousins $R$-band (right ordinate) as taken and sometimes
recalibrated from Stanek et al 2001a and 2001b, Price et al. 2001,
Orosz 2001, Masetti et al. 2001a, Oksanen et al. 2001, Stanek \& Falco 2001,
Valentini et al. 2001, Watanabe et al. 2001, Veillet 2001 and Garnavich
et al. 2001b, together with the broken power-law fitted to these data
(dashed curve) with a break time of $41.5\times10^3$~s and decay indices
$\alpha$ of $0.60$ before the break and $1.31$ afterwards (Masetti et al.
2001b). The same
curve is also shown shifted to the level of the MECS data and taking $\alpha$ before
the break to be 0.85 as would be expected for adiabatic cooling. The dash-dotted
curve shows the same function except that $\alpha=0.60$ before the
break (to illustrate the case of radiative cooling).
The dotted curve shows the best-fit power law function for the LECS/MECS data.
All broken power laws fit the late WFC data better than the unbroken power
law. However, for all power laws, the CXO data point (bottom right-hand corner)
is a factor of 2 too faint with respect to the extrapolation of those.
\label{fignfilc}
}
  \end{center}
\end{figure}

Optical photometry revealed an early break in the light curve (Holland et al.
2001), see figure~\ref{fignfilc}. Masetti et al. (2001b) carried out UBVRIJK
photometry and determined that the break is achromatic. When fitted with a
smoothed
broken power law as
defined by Beuermann et al. (1999), they find in the $R$-band a break time of
$0.48\pm0.02$~d and decay indices of $0.60\pm0.03$ and $1.31\pm0.03$ before
and after the break. The B, V and I light curves are convincingly consistent
with the broken power law; the J and K band are not sufficiently sampled
to make a judgment. The slope change occurred relatively quickly, there
is no difference with an unbroken power law from 1 day after the GRB on. The
latest published $R$-band magnitude was obtained 25 days after the burst 
(Garnavich, Quinn \& Stanek 2001b) and
is still consistent with the function parameterized by Masetti et al. (2001b).

Thanks to the long WFC coverage after the burst (3.5 ~hr) and the brightness
of the prompt and afterglow emission, we are able to set
significant constraints on the early X-ray afterglow. While the NFI data are
satisfactorily and accurately described by an unbroken power-law decay, the
last two WFC data points lie significantly below the extrapolation of that
power law (see figure~\ref{fignfilc}). The one but last data point is $5\pm1$
times fainter and the last data point, a $3\sigma$ upper limit, is $1.5\pm0.1$
times fainter. Thus, a break is prescribed in X-rays. We have fitted the
smoothed broken power law function found by Masetti et al. (2001b) to the NFI
data, leaving free only the normalization and employing $\alpha=0.60+0.25=0.85$
before the break as would be expected for adiabatic cooling and for a
cooling frequency between the X-ray and optical bands, and $\alpha=0.60$
as would be expected for radiative cooling (Sari, Piran \& Narayan 1998),
see figure~\ref{fignfilc}. Given that the afterglow is expected to
begin during the main event (e.g., Frontera et al. 2000), these functions
are more consistent with the late WFC data than the unbroken power
law. We conclude that the X-ray data support the achromatic nature of the
break seen in the optical.
This is only the second time that a light curve break in a GRB afterglow is
seen in X-ray data (Kuulkers et al. 1999, Pian et al. 2001).
However, we note that the first 3 NFI data points are up to 2$\sigma$
above the broken power law functions. Furthermore, the preliminary X-ray
flux as measured with CXO at nine days after the burst by Harrison et al.
(2001b, see figure~\ref{fignfilc}) lies 48$\pm$6\% below the extrapolated
unbroken power law and is similarly inconsistent with the broken power law.
Therefore, the suggestion of the achromatic nature of the break is not as
straightforward as one might hope.

\begin{figure}[t]
  \begin{center}
    \leavevmode
\epsfxsize=10cm
\epsfbox{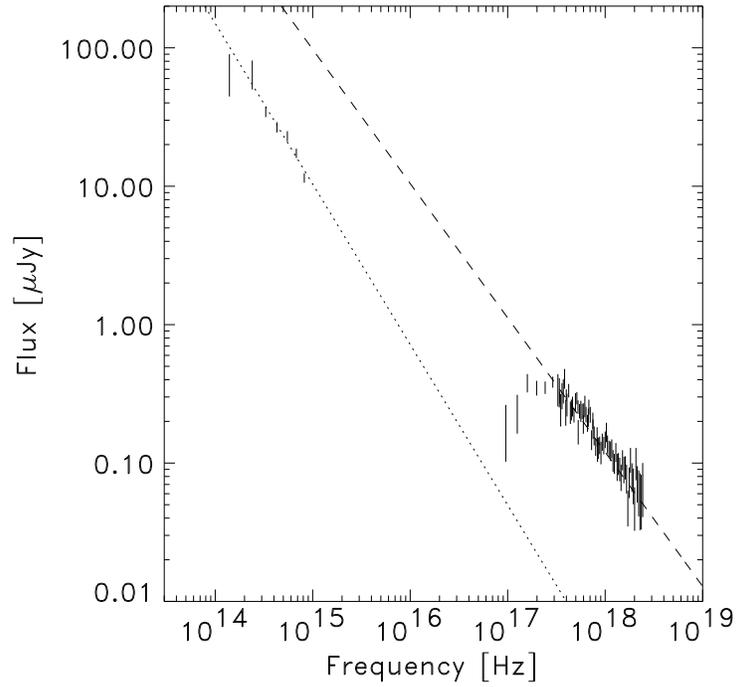}
  \caption{Broadband spectrum from UBVRIJK photometry and X-ray spectrum at
0.97~d after
the burst. The X-ray spectrum is for the whole NFI observation, but scaled
to the flux level as observed at 0.97~d. No correction for absorption was
carried out. The power law functions fitted to each data set are shown
(dotted line for optical data with spectral index 1.16 [Masetti et al. 2001]
and dashed line for X-ray data with spectral index 0.97)
\label{figbroad}
}
  \end{center}
\end{figure}

In figure~\ref{figbroad}, the optical photometry (Masetti et al. 2001b) and
X-ray spectrum are shown for 0.97~d after the burst. Both data sets
individually have similar spectral indices, but they do not match each other's
extrapolations. The optical data are about 2.5 mag too faint or the X-ray
data a factor of about 30 too bright. This cannot be explained by standard
extinction; possibly other
extinction laws apply, as was already noted from absorption bands in the
optical spectrum by Jha et al. (2001b) and Masetti et al. (2001b).
Alternatively, the jump in the spectrum may be related to an important
contribution to the X-ray spectrum of inverse Compton scattering
while the optical spectrum is dominated by synchroton emission (e.g.,
Sari \& Esin 2001). This was recently suggested in a similar observation
of another burst (Harrison et al. 2001a). Whatever the cause, the similar
spectral indices
indicate that there is no spectral break between the optical and X-ray bands.
This is also consistent with the decay indices at these two wavelength
regimes being equal (in fact, the X-ray decay index is the most accurate
obtained for any GRB afterglow thus far, and has a similarity to the optical
decay index of $2\pm4$\% which is highest degree of agreement of all
afterglows). This is in
contrast to some other GRBs where the spectral break at the cooling frequency
is between the optical and X-rays.

Currently, it is thought that two mechanisms may be responsible
for achromatic breaks in GRB afterglows. In the first, the fireball 
has a collimated expansion (jet) and the break occurs when the bulk Lorentz
factor becomes smaller than the inverse of the opening angle
of the jet (Kulkarni et al. 1999; Sari, Piran \& Halpern 1999; Livio \&
Waxman 2000). Measurement of the break time directly provides the opening
angle and the total amount of radiated energy. After the break, $\alpha$ and
$\beta$ are related to the power-law index of the Lorentz-factor distribution
of the electrons $p$ through $\alpha=p$ and $\beta=p/2$. The second mechanism
concerns the deceleration of the fireball into the non-relativistic (NR)
domain in
a dense circumburst medium (Dai \& Lu 1999); the break occurs when the rest
mass of the material swept up by the fireball equals the initial fireball
energy. In this mechanism, $\alpha=(3p-4)/2$ and $\beta=p/2$. 
For the jet interpretation of the break, the two values for $p$
determined from $\alpha$ and $\beta$ are $1.33\pm0.03$ and
$1.93\pm0.10$. The difference is 5.7$\sigma$ and the weighted mean
1.38. For the NR effect, the $\alpha$ and $\beta$-derived values for $p$ are
$2.22\pm0.03$ and $1.93\pm0.10$. Here the difference is 2.8$\sigma$ and the
weighted mean 2.20. Clearly, the NR interpretation is favored over the jet
interpretation because the difference between the two $p$ values is consistent
with zero and the mean value of $p$ is not smaller than 2 which would have
implied an unbound energy estimate for the electrons. A similar conclusion
was reached from optical observations alone by Masetti et al. (2001).
For the fireball to
become non-relativistic within 1~d after the burst, theory predicts (Blandford
\& McKee 1976) that, for an energy of $\sim10^{53}$~erg (see below), a density
for the circumburst medium of $\sim10^6$~cm$^{-3}$ is required. The afterglow
shows absorption equivalent to a
redshift-corrected column density of $\sim10^{22}$~cm$^{-2}$. Combining these
two numbers implies a size of $\sim10^{16}$~cm, or $10^5$~lt-sec for the dense
part of the circumburst medium. This is large enough for the presumed fireball
to be able to slow down to non-relativistic speeds.

Given $E_{\rm iso}(\gamma)=7.8\times10^{53}$~ergs (equivalent to 0.4~M$_\odot$
rest mass), it is likely that beaming is important in the initial stages of the
burst, because no presently considered progenitor model is consistent
with such a large energy (for similar arguments as applied to GRB~990123, see
Kulkarni et al. 1999). However, there is no evidence in the data for a jet
presence: one would have expected a steeper decay index that should also be
larger than 2 before the NR-induced break (e.g., Livio \& Waxman 2000; see
also Piro et al. 2001b where such behavior is shown for GRB~000926). Instead,
the decay index is much shallower. The only way to make a jet 'invisible'
like that is to require that the jet-induced
light-curve break happens simultaneous with the NR-induced one. From the
formalism of Sari et al. (1999), and assuming an efficiency to
convert energy to radiation of 1 and a density of $\sim10^6$~cm$^{-3}$, it
follows that the opening angle $\theta_j$ is of order 15$^{\rm o}$. The
beaming-corrected energy is then $3\times10^{52}$~ergs. This implies an
unprecedented high energy content. However, this is uncertain,
because it depends on whether early breaks in GRB~990123 and
GRB~000131 are truly due to a jet (e.g., Dai \& Lu 1999 for GRB~990123).

It should be noted that there is a consideration which favors a low-density
interpretation of the observed light curve break. The jet opening angle
for a mean circumburst density of 0.1~cm$^{-3}$ is
$\theta_j$=2\degr. Thus, the beaming-corrected energy output
$E_\gamma$ is $(5.7\pm0.2)\times10^{50}$~erg. Frail et al. (2001)
similarly apply a correction for beaming in 15 GRBs for which this is
possible and find that $E_\gamma$ has a range of
0.234-1.80$\times10^{51}$~erg with a mean of $5\times10^{50}$~erg. This
dynamic range is three times smaller than that of the uncorrected energies,
which is compelling evidence that beaming is a dominant mechanism responsible
for light curve breaks and that densities are low. For GRB~010222, $E_\gamma$
happens to be very close to the mean value for these 10 bursts (i.e., within
14\%).

\section{Conclusion}

So far, GRB~010222 is the third most energetic GRB for which a
distance has been determined. We measured the 2-10 keV light curve during
the first 65~hr after the burst, except for a gap between 4 and 8~hr,
and find that there is a break which is consistent with a break
in the optical band. The jet-interpretation of this break is not
straightforward,
and an interpretation in the form of a quick brake of the fireball into
a dense circumburst medium appears to be more consistent with the data.
If this is indeed true, GRB~010222 may very well be the most energetic
burst thus far detected.


\acknowledgements
We thank Elena Pian for her contribution to this work. Furthermore,
we are grateful to the staff of the BeppoSAX Scientific Operation Center,
the Mission Planning Team and the Science Data Center for their support in
obtaining and processing the data. JZ and EK acknowledge financial support
from the Netherlands Organization for Scientific Research (NWO).
The BeppoSAX satellite is a joint Italian and Dutch program.



\newpage
\begin{table}[t]
\caption[]{Basic parameters of the prompt emission.\label{table1}}
\begin{tabular}{llll}
\hline\hline
Parameter & unit & bandpass & value \\
\hline
Duration  & s & 2-28 keV   & $\approx$280 \\
          &   & 40-700 keV & $\approx$170 \\
Peak flux & erg~s$^{-1}$cm$^{-2}$& 2-10 keV   & $(2.1\pm0.2)\times10^{-7}$ \\
          &   & 2-28 keV   & $(4.6\pm0.5)\times10^{-7}$ \\
          &   & 40-700 keV & $(8.6\pm0.2)\times10^{-6}$ \\
          &   & 50-300 keV & $(3.6\pm0.4)\times10^{-6}$ \\
          & phot~s$^{-1}$cm$^{-2}$ & 50-300 keV & $19\pm2$  \\
Fluence & erg~cm$^{-2}$   & 2-10 keV   & $(1.03\pm0.03)\times10^{-5}$ \\
          &   & 2-28 keV   & $(2.20\pm0.06)\times10^{-5}$ \\
          &   & 40-700 keV & $(9.25\pm0.28)\times10^{-5}$ \\
          &   & 2-700 keV  & $(1.20\pm0.03)\times10^{-4}$ \\
          &   & 50-300 keV & $(4.88\pm0.13)\times10^{-5}$ \\
\hline\hline
\end{tabular}
\end{table}

\end{document}